\documentstyle[aps,twocolumn]{revtex}  
                                  
\begin{document}
\title{Stability of periodic orbits controlled by time-delay feedback}
\author{M.~E.~Bleich and J.~E.~S.~Socolar}
\address{Department of Physics and Center for Nonlinear and Complex Systems,
	 Duke University, Durham, NC 27708}
\maketitle

\begin{abstract}
Extended time-delay auto-synchronization (ETDAS) is a promising technique
for stabilizing unstable periodic orbits in low-dimensional dynamical systems. 
The technique involves continuous feedback of signals delayed by multiples
of the orbit's period in a manner that is especially well-suited
for fast systems and optical implementation.
We show how to analyze the stability of a given implementation
of ETDAS without explicit integration of time-delay equations.
To illustrate the method and point out some nontrivial
features of ETDAS, 
we obtain the domain of control for a period-one orbit
of the driven, damped pendulum.
\end{abstract}

\section{Introduction}

The prospect of ``controlling chaos'' has generated great interest
among physicists over the past several years.
As first pointed out by Ott, Grebogi, and Yorke, \cite{OGY} 
the existence of many periodic orbits embedded in strange
attractors raises the possibility of using very small
control signals to obtain various types of regular
behavior from intrinsically chaotic systems.
This fact has implications both for engineering nonlinear systems
in which chaotic fluctuations occur but are undesirable
and for the understanding
of biological systems in which underlying nonlinear dynamical
systems are regulated in ways that are at present poorly understood.

The initial problem that must be faced in developing
a control mechanism requiring only small externally imposed 
perturbations to the system is to design a feedback scheme
that allows an unstable periodic orbit (UPO) to be stabilized.
Recently several techniques have been introduced for 
accomplishing this using feedback signals that actually vanish 
(in the absence of noise)
when the system is on the desired orbit.\cite{otherOGY}
One such technique, sometimes called
``time-delayed autosynchronization'' (TDAS) \cite{Pyragas1},
involves a control signal formed from the difference
between the current state of the system and the state of the system
delayed by one period of the UPO.
One of us (JESS) and coworkers have showed how to efficiently
reuse information generated further in the past \cite{SSG},
using a technique
called ``extended time-delayed autosynchronization'' (ETDAS).    
In the extended scheme, discussed in detail below,
the control signal consists of a particular linear
combination of signals from the system delayed by 
integer multiples of the UPO's period.
TDAS is a special, limiting case of ETDAS.

ETDAS has several features of practical interest. \cite{Pyragas1,SSG}
First, the use of a time-delay in the feedback loop 
eliminates the need for explicitly determining any information 
about the underlying dynamics other
than the period of the desired orbit.
Second, it can be implemented using a continuous feedback loop
and hence can be applied to stabilize oscillations that are
too fast to be handled by standard techniques based on
measurements of the system on a surface of section.
Finally, ETDAS provides a natural choice for controlling chaos using all-optical
methods, as the feedback signal corresponds precisely
to the reflected signal from a Fabry-Perot interferometer with
properly adjusted cavity length. \cite{Lu,GSCS}

This paper concerns the stability analysis 
of ETDAS in continuous systems
for which the dynamical equations are known.
We note that versions of ETDAS that apply to discrete maps
can be treated analytically and it is known that
ETDAS can stabilize orbits that are uncontrollable using TDAS.\cite{SSG,abed1}
For continuous systems,
both numerical results \cite{Pyragas1,Pyragas3}
and experiments \cite{SSG,GSCS,Pyragas2,Bielawski}
have shown that in order for ETDAS to be successful
the feedback gain must lie within a finite, and often narrow, range.
As the UPO is modified by changes in a bifurcation parameter
this range of successful feedback gain will in general shift. 
In the space of the feedback gain and a bifurcation parameter,
the area for which ETDAS can be successfully applied is 
known as the domain of control.
Here we show how the domain of control can be obtained, and
as an example, we find such a domain for the nonlinear pendulum.
Though the domain of control could also be mapped using
direct integration of the time-delay differential equations
governing the controlled system, such a procedure may
encounter difficulties associated with the accuracy of
the integrator over long times and the choice of initial conditions.
Our technique avoids both of these difficulties, requiring
only the integration of equations with no time delay
over one period of the UPO.

The results from the pendulum demonstrate the
effectiveness of ETDAS in controlling highly unstable orbits that are
impossible to stabilize with TDAS.
The analysis also shows that the domain of control is
surprisingly complex and depends strongly on the particular
choices of the accessible control parameter and 
the measured signal used to generate the feedback. 
The domains of control have several features that
clearly distinguish the application of continuous control
of a period-one orbit from the control of a fixed point of
a discrete map.

Two items that may be important in experimental implementations
of ETDAS but are not treated here are the effects of noise
and the unavoidable small time lag between the measurement of
the system and the application of feedback.
Preliminary numerical investigations and experiments on fast
diode resonators \cite{SG} indicate that low noise levels
and small time lags shift the boundaries of the domain of control
only by small amounts.
The results we obtain for the idealized situation do indeed
give a useful guide to the phenomenology of real systems.

\section{Stability Analysis}

A general form for an $N$-dimensional system governed by 
ordinary differential equations 
and controlled via variations in an accessible parameter $\kappa$ is
\begin{eqnarray}
\label{sys}
 {\bf \dot{x}}(t) & = & {\bf f}\left( {\bf x}\left( t \right),t; \kappa \right),\\
           \kappa & = & \kappa_0 + \epsilon(t), \nonumber
\end{eqnarray}
where ${\bf x}(t)$ and ${\bf f}$ are $N$-dimensional vectors
and $\epsilon(t)$ is a small control signal. 
In ETDAS, the control signal is given by
\begin{equation}
\label{epsilon}
 \epsilon (t) = \gamma [\xi(t) - (1-R)\sum_{k=1}^{\infty} R^{k-1} \xi(t-k\tau)],
\end{equation}
where $\xi$ is some measured
component of ${\bf x}$, $\tau$ is the period of the desired UPO,
and $R \in [0,1)$ is a real parameter.
The case $R=0$ corresponds to TDAS.
Note that when control is successful,
the system is synchronized with its own past behavior, so
$\xi(t-k\tau) = \xi(t)$ for all $k$ and $\epsilon(t)$ vanishes.
Thus to analyze the stability of the controlled system with respect
to small perturbations, it is sufficient to consider a linearized form
of Eqn.~(\ref{sys}) in which both $\epsilon(t)$ and the 
deviations from the UPO are considered small.
Letting ${\bf x}_0(t)$ be the UPO and 
${\bf y}(t)$ be the deviation from this orbit,
${\bf y}(t) \equiv {\bf x}(t) - {\bf x}_0(t)$, we have
\begin{equation} 
\label{eq:1}
 {\bf \dot{y}}(t) = {\bf J}(t)\cdot {\bf y}(t) + 
 \epsilon(t) \frac{\partial {\bf f}}{\partial \kappa},
\end{equation}
where ${\bf J}\left( {\bf x}_0(t) \right) \equiv 
	\frac{\partial {\bf f}}{\partial {\bf x}} 
	\mid _{{\bf x}_0(t),\kappa_0}$
is the Jacobian of the uncontrolled system.
Eqn.~(\ref{eq:1}), with $\epsilon$ given by Eqn.~(\ref{epsilon}),
can also be written as
\begin{eqnarray} 
\label{etdas}
 {\bf \dot{y}}(t)  & = & {\bf J}(t)\cdot {\bf y}(t) \nonumber \\ 
& + & \gamma {\bf M}(t) \cdot \left[ {\bf y}(t) - (1-R)\sum_{k=1}^{\infty}
 R^{k-1} {\bf y}(t-k\tau) \right],
\end{eqnarray}
where  ${\bf M}({\bf x}_0(t)) \equiv
	(\frac{\partial {\bf f}}{\partial \kappa}
	\mid _{{\bf x}_0(t),\kappa_0} ) \otimes \hat{\bf n}$,
is an $N \times N$ dyadic which contains all information about how the 
control is applied to the system;  
$\hat{\bf n}$ is a constant unit vector that
determines the component of ${\bf x}$
that enters the control signal via $\xi = \hat{\bf n} \cdot {\bf x}$,
and $\frac{\partial {\bf f}}{\partial \kappa}$
describes the effect on the dynamics
of small changes of the control parameter $\kappa$.
Eqn.~(\ref{etdas}) applies
to both periodically driven and autonomous systems.
For periodically driven systems, 
the period $\tau$ is equal to the period of the drive 
(or an integer multiple of it); 
for autonomous systems $\tau$ is not known a priori,
but it can be repeatedly adjusted until control is achieved. 
(One scheme for making the adjustment 
is discussed by Kittel et al. \cite{Pyragas3})
Our goal is to investigate the stability of the trivial
${\bf y}=0$ solution of Eqn.~(\ref{etdas}),
which corresponds to the system remaining on the UPO.

Eqn.~(\ref{etdas}) is a special case of a 
form that has been treated
in the mathematics literature. \cite{Hale}
It can be written as
\begin{equation} 
\label{eq:2}
 {\bf \dot{y}}(t) = \sum_{n=0}^{\infty} 
 {\bf A}_{n}(t) \cdot {\bf  y}(t-n\tau),
\end{equation}
where ${\bf y}(t)$ is an $N$-dimensional vector and
each ${\bf A}_{n}(t)$ is an $N \times N$ matrix
with elements that are periodic in time with period $\tau$.
Specifically, we have
\begin{equation} 
\label{AJM}
 {\bf A}_n(t) = \left\{ 
		\begin{array}{ll}
		 {\bf J}(t) + \gamma {\bf M}(t) & \mbox{for $n=0$} \\
		 - \gamma (1-R)R^{n-1}{\bf M}(t) & \mbox{for $n=1,2,3,... .$}
		\end{array}
		\right.
\end{equation}
Notice that ${\bf J}$ and ${\bf M}$ (and consequently ${\bf A}_n$)
are periodic with period $\tau$ by virtue of the fact that they
are evaluated along the UPO.

The general approach to the linear stability of integer
time-delay differential equations with periodically varying
coefficients has been discussed by Hale and Verduyn Lunel \cite{Hale}.
(The one-dimensional case has also been
addressed by Ortega \cite{Ortega}.)
We briefly outline here a derivation of their central result,
which then will be used as the basis for an efficient numerical
technique for mapping the domain of stability in the appropriate
parameter space.

By virtue of the linearity of Eqn.~(\ref{eq:2}),
a general solution can be composed from a sum of periodic modes
with exponential envelopes:
\begin{equation} 
\label{sub}
 {\bf y}_k (t) = {\bf p}_k (t) {\rm exp}(\lambda_k t /\tau), 
\end{equation}
where ${\bf p}_k (t+\tau) = {\bf p}_k (t)$
is an $N$-dimensional vector and
$\lambda$ is a complex number.  
For one such mode, one obtains from Eqn.~(\ref{eq:2})
(dropping the subscript $k$)
\begin{equation} 
\label{periodicRE}
 \dot{\bf p}(t) = ( \sum_{n=0}^{\infty} e^{-n\lambda}{\bf A}_{n}(t)
                    - \frac{\lambda}{\tau} ) \cdot {\bf p}(t).
\end{equation}
Equivalently we can write
\begin{equation}
\label{poft}
 {\bf p}(t) = e^{-\lambda t/\tau} {\bf U}(t) \cdot {\bf p}(0),
\end{equation}
where the matrix ${\bf U}(t)$ is the solution of the equation
\begin{equation} 
\label{Ueq}
 {\bf \dot{U}}(t) = 
 \sum_{n=0}^{\infty} e^{-n\lambda}{\bf A}_{n}(t) \cdot {\bf U}(t)
\end{equation}
with ${\bf U}(0) = 1\!\!1$.
Defining the Floquet multiplier $\mu \equiv e^\lambda$ a
formal solution for ${\bf U}(\tau)$ can be written as
\begin{equation}
\label{top}
 {\bf U}(\tau) =  T\left[ e^{\int_{0}^{\tau}dt \sum_{n=0}^{\infty}
 \mu^{-n}{\bf A}_{n}(t)}\right],
\end{equation}
where $T[ \cdots ]$ indicates the time-ordered product. 
The time-ordered exponential is 
simply a compact notation \cite{time-ordering}
used to emphasize the way 
in which ${\bf U}(\tau)$ depends on $\mu$.
In general, ${\bf U}(\tau)$ must be obtained by direct
numerical integration of Eqn.~(\ref{Ueq}).

In order for
${\bf p}(t)$ to be periodic,
Eqn.~(\ref{poft}) implies that
${\bf p}(0)$ must satisfy the equation
\begin{equation}
 \left(\mu^{-1}{\bf U}(\tau) - {\bf U}(0)\right)\cdot {\bf p}(0) = 0,
\end{equation} 
which in turn requires the vanishing of the determinant of
$\left(\mu^{-1}{\bf U}(\tau) - {\bf U}(0)\right)$.
Substituting for the ${\bf U}'s$ one obtains 
a modified eigenvalue equation for $\mu$:
\begin{equation}
\label{fldef}
 \left| \mu^{-1} T\left[ e^{\int_{0}^{\tau}dt \sum_{n=0}^{\infty}
 \mu^{-n}{\bf A}_{n}(t)}\right] - 1\!\!1  \right| = 0.
\end{equation}
Eqn.~(\ref{sub}) shows that
the trivial solution of Eqn.~(\ref{eq:2}) is asymptotically
stable if and only if all $\mu$ which satisfy Eqn.~(\ref{fldef})
also satisfy $\left| \mu \right| < 1$.
This is the central result advertised above.

Inserting Eqn.~(\ref{AJM}) into Eqn.~(\ref{fldef}) and performing the
geometric sum over coefficients of ${\bf M}$, we have
\begin{equation}
\label{def}
 \left| \mu^{-1} T\left[ e^{\int_{0}^{\tau}dt\left( {\bf J}(t) +
 \gamma \frac{1-\mu^{-1}}{1-\mu^{-1} R} {\bf M}(t) \right) } \right] 
 - 1\!\!1 \right| =0,
\end{equation}
the defining relation for Floquet multipliers of systems
under ETDAS control.
The system is linearly  (or locally asymptotically) stable if and only if
$\left| \mu^{-1} \right| > 1$
for all $\mu$ satisfying Eqn.~(\ref{def}).
\cite{roots}.
For $R<1$, the determinant on the
left hand side of Eqn.~(\ref{def}), which
will be denoted $g(\mu^{-1})$, 
has no poles inside the unit circle.
Hence, by a well-known theorem in complex analysis,
the number of roots of $g(\mu^{-1})$ with
$\left| \mu^{-1} \right| <1$ is equal to 
the number of times the path traced by $g(\mu^{-1})$ 
winds around the origin 
as $\mu^{-1}$ is varied one full time around the unit circle.
\cite{Henrici}
The condition for linear stability of the controlled system
is that this winding number, 
which will be denoted ${\cal N}$, vanishes. \cite{autonomous-note}

For generic ${\bf J}(t)$ and ${\bf M}(t)$, 
the time-ordered integral discussed above cannot be obtained
in closed form,
and so must be computed numerically by explicitly
integrating Eqn.~(\ref{Ueq}). \cite{pre}
Note, however, that this is an ordinary integration
which does not involve any time-delayed quantities.
Thus we have avoided the integration of a 
time-delay differential equation, 
for which the issue of how to choose initial 
conditions can be rather delicate.

\section{Numerical Procedures}

In a typical situation, the system parameters that
can be externally adjusted are largely dictated
by physical principles and practical considerations.
The problem is therefore to determine whether control
can be achieved for a given designation of the control parameter $\kappa$.
In general, we may expect the success of ETDAS in controlling
highly unstable orbits to depend upon the choice of which
system variable is used to construct the feedback signal;
i.e., the choice of $\hat{\bf n}$.
We wish to determine the domain of values of
$\hat{\bf n}$, $\gamma$, and $R$ for which ETDAS is successful
for a given $\kappa$.
If the dynamical equations governing the system are known,
the results can provide direct guidance in selecting
appropriate parameters for operating the controlled system.
The exercise of computing the domain of control for a simple
model is also useful in that it reveals qualitative features
that should be kept in mind when trying to find a stable
regime in a system for which the equations are {\em not} known.

In general, the function ${\bf f}$ in Eqn.~(\ref{sys}) depends on
a ``bifurcation parameter'', which we denote $r$.
As $r$ is varied, the properties of a given UPO change,
so the stability must be considered separately for different $r$.
We choose to map the domains of control 
in the plane of the bifurcation parameter, $r$,
and the feedback gain, $\gamma$, for several discrete choices of
$\hat{\bf n}$ and $R$.

Calculation of the domain of control for a system given by Eqn. 
(\ref{sys}) and particular choices of $\kappa$, $\hat{\bf n}$, and $R$
involves three distinct numerical tasks:
\begin{enumerate}
\item the determination of
 the desired periodic orbits of the uncontrolled system;
\item the calculation of $g(\mu^{-1})$ for a given $\mu$ on the unit circle
 and given values of $r$ and $\gamma$; and
\item the evaluation of the winding number, ${\cal N}$,
 which determines the number of unstable modes.
\end{enumerate}
The first is easily accomplished using a standard Newton's method.
Using the period of the resulting solution and a point on the orbit, 
the second task requires straightforward simultaneous integration of
the uncontrolled dynamical equations and
the set of first-order ordinary differential equations (Eqn.~(\ref{Ueq}))
that determine ${\bf U}(\tau)$.
In the cases we have studied,
a fifth order adaptive step-size
Runga-Kutta method has proven satisfactory.
Finally, ${\cal N}$ can be determined 
by evaluating $g$ for a sequence of
sufficiently closely spaced $\mu$'s 
around the unit circle and considering the 
sequence of values of ${\rm arg} g(\mu^{-1})$.
The necessary number of points in the sequence depends
upon the proximity of roots of $g$ to the unit circle.

With these tools in hand, the boundary of the domain
of control may be located by the following method.
First, a single point on the boundary must be determined.
If the orbit of the uncontrolled system
becomes unstable when $r=r_c$, a point
on the boundary of the domain of control 
can always be found at $(r, \gamma) = (r_c, 0)$
since for $\gamma=0$ the controlled system is identical to the
system without feedback control.
An entire boundary may then be traced by
changing $r$ in small increments, each time searching
over a small interval in $\gamma$ to locate the boundary.
The boundary of the domain of control is identified
as the point where the winding number jumps 
from zero to a positive integer, 
signalling the entry of a root into the unit circle. \cite{autonomous-note}

Additional work may be necessary in order to find all the islands
of stability, since
the domain of control may not be simply connected.
We have encountered this situation only in the case where
a given periodic orbit is unstable only over a finite
interval in the parameter $r$ with endpoints $r_1$ and $r_2$.
In this case, there are two separate regions of stability,
each of which can be found starting from $r=r_{1,2}$, $\gamma=0$.
(See the example discussed below.)
We have not been able to rigorously rule out the occurrence 
of more complicated situations in
which a domain has a hole in it or an island of stability
exists that does not merge into an intrinsically stable regime.
We have tested for this by scanning through large intervals of
$\gamma$ for selected values of $r$ in two different systems,
the pendulum discussed below and a diode-resonator circuit that
will be the subject of a future publication \cite{SG},
and have yet to observe such behavior.

As a check of the above analysis and numerical procedure,
selected sections of the stability domains presented 
below were checked with explicit integration of the 
time-delayed differential equations given by Eqn.~(\ref{sys}).  
Using a fourth order Adams-Bashforth-Moulton
algorithm modified for time-delayed equations, we found
complete agreement between the two methods to the
expected degree of accuracy.
We emphasize, however, that our procedure is
more reliable and significantly easier to automate.

\section{An Example: The Nonlinear Pendulum}

To demonstrate the utility of the stability analysis described above
and begin to investigate the structure of domains of control,
we calculate stability domains for the damped driven
nonlinear pendulum:
\begin{eqnarray}
\dot{x_1} & = & x_2 \\
\dot{x_2} & = & -\nu x_2 - \sin x_1 + F \cos (\omega t).
\end{eqnarray}
Here $x_1$ is the angle of the pendulum and $x_2$
is its angular velocity;  
$\nu$, $F$, and $\omega$ describe the damping, drive amplitude and 
drive frequency, respectively.  
We fix $\nu=1/2$ and $\omega= 2 \pi /10$, 
and vary $F$ as the bifurcation parameter.

We choose to study a family of period-one orbits which are
unstable between $F \simeq 0.987$, and $F \simeq 2.046$.
These orbits and their largest Floquet multipliers
are shown in Fig.~1.
Note that some of these orbits are highly unstable,
having Floquet multipliers as high as 30.
The explicit demonstration (below) 
that periodic orbits with multipliers this large
can be stabilized 
using continuous time-delay techniques is a new result,
though a plausibility argument has been given
based on the behavior of discrete maps. \cite{SSG}

ETDAS requires that we measure some
accessible quantities in the system and feed back the control signal
through small modifications of an accessible parameter.
We assume that both the position
and velocity of the pendulum are measurable, and that 
small changes can be made to the amplitude of the drive, 
$F \rightarrow F + \epsilon (t)$.
(Note that $F$ is being used here in a dual role,
both as the bifurcation parameter, $r$, and the
accessible control parameter, $\kappa$.)
The control signal is generated from a linear combination
of deviations in position and velocity.
In the notation of Eqn.~(\ref{etdas}), we have
${\bf \hat{n}} = (\sin \phi, \cos \phi)$, 
where $-\pi/2 \leq \phi \le  \pi/2$
is a parameter that can be chosen to optimize the domain of control.
The range $-\pi/2 \leq \phi \leq \pi/2$
is sufficient to describe the entire space of possibilities
since the case $\phi + \pi$ is equivalent to $\phi$
with the sign of the feedback gain reversed.
Note that $\phi = 0$ corresponds to measuring velocity only, 
and $\phi = \pi/2$ corresponds to measuring position only.
The control matrix ${\bf M}$ is given by
\begin{equation}
{\bf M}(t) = F\cos (\omega t) \left( \begin{array}{cc}
		0 & 0 \\
		\sin \phi & \cos \phi
		\end{array}
	    \right).
\end{equation}

Our numerical implementation of the program outlined above was
straightforward.
The most significant source of potential errors is the possibility
that a root of $g(\mu^{-1})$ crosses into the unit circle
but remains extremely close to the boundary and is not picked up
in the winding number calculation.
This problem can be solved to any desired accuracy by choosing
sufficiently many points around the unit circle in evaluating ${\cal N}$
(or by using adaptive step-size methods).
For the present case, 500 equally spaced points were used.
This large number was necessary, however, only to obtain accurate
results in regimes with very narrow features.
In general, the necessary spacing between points is determined
by the rate at which the first root enters the unit circle as
$\gamma$ is varied across the stability boundary.

The dependence on $R$ of the domain of control is shown in Fig.~2
for $\phi = -\pi/8$ which is representative of all values of $\phi$
that we investigated.
A key point that is evident here is that large values of
$R$ are necessary in order to
control the highly unstable periodic orbits.
In addition,
the domain of control strongly depends on the choice of $\phi$,
as might be expected by analogy with the situation
for proportional feedback control of a stationary fixed point.
Fig.~3 shows the domain of control for $R=0.95$
and several choices of $\phi$.
The domain extends across the entire
unstable region only for $\phi$ within a relatively narrow range near
$\phi = 0$, indicating the relative merit of measuring the
velocity of the pendulum for our example in which the control
signal is applied to the amplitude of the drive.
Even for $R$ close to $1$, it is not always possible to
control highly unstable orbits for arbitrary $\phi$.

The results depicted in Fig.~3 illustrate the
complexity of the domain of control.
The sharp features and the existence of reentrant
behavior for varying $\gamma$ indicate that
intuition about the qualitative shapes of these domains
may be highly misleading.
An interesting example of the counterintuitive phenomena
that can occur 
can be seen in the
case $\phi = \pi/8$.
As shown in Fig.~4,
there is a region for which both $\gamma$ and $-\gamma$
successfully stabilize an UPO.
In this region, one has
an orbit that is unstable in the absence feedback.
Naively, one would expect that if a given form of linear
feedback resulted in stabilization of the orbit, then
reversing the sign of the feedback gain would make the orbit
even more highly unstable.
In the present case, however, the variation of $\bf J$ 
around the orbit allows the inverted feedback gain
to be equally effective in stabilizing the orbit.
This can never occur for the case of linear control of
a stationary fixed point in a continuous system or
a discrete map.

The most important point here is that 
the experimental determination
of the ETDAS domain of control for a given system
should be guided as much as possible by calculations
on model equations.
A coarse scan of parameter space guided by
``intuitively reasonable'' ideas concerning
the possible structure of these domains may well
miss regimes of practical interest.

\section{Conclusions}

The primary purpose of this paper is to demonstrate
the proper technique for analyzing the stability of
systems controlled using ETDAS.
Generic systems do not permit a complete analytical
solution, even for the linear stability problem.
The logic of our approach does, however, allow the
development of a clean numerical technique.
In particular, it avoids the need for integrating
delay-differential equations directly and thereby
avoids the difficulties associated with guaranteeing
that a chosen initial condition lies in the appropriate
basin of attraction.

The analysis of an orbit of the nonlinear pendulum
confirms the fact that ETDAS can work well in continuous
systems and also clearly illustrates the complexity of
this linear stability problem.
We have also studied a set of equations describing
a fast diode resonator and compared our results
with experiments.
Details will be published together 
with the experimental results.
Here we note only that the effects of noise and
a time lag in the feedback loop outside the
recursively used delay line \cite{SG}
do alter the domains of control slightly.
Inclusion of these effects and extensions
of the technique to spatially extended systems
will be addressed in future studies.

We thank D. Gauthier, H. Greenside, J. Sethna, and D. Sukow 
for helpful conversations
and D. Gauthier for a critical reading of the manuscript.
This work was supported by NSF Grant DMR-9412416.


\begin{references}
\bibitem{OGY}  E.~Ott, C.~Grebogi, and J.~Yorke, 
	Phys. Rev. Lett. {\bf 64}, 1196 (1990).
\bibitem{otherOGY}  For a review of techniques that stabilize UPOs see 
	T.~Shinbrot, C.~Grebogi, E.~Ott, and J.~A.~Yorke, 
	Nature {\bf 363}, 411 (1993).
\bibitem{Pyragas1}  K.~Pyragas, 
	Phys. Lett. A~{\bf 170}, 421 (1992).
\bibitem{SSG}  J.~E.~S.~Socolar, D.~W.~Sukow, and D.~J.~Gauthier,
	Phys. Rev. E~{\bf 50}, 3245 (1994).
\bibitem{Lu}  W.~Lu and R.~G.~Harrison,
	Optics Comm.~{\bf 109}, 457 (1994).
\bibitem{GSCS}  D.~J.~Gauthier,  D.~W.~Sukow, H.~M.~Concannon, and J.~E.~S.~Socolar, 
	Phys. Rev. E~{\bf50}, 2343 (1994).
\bibitem{abed1} E.~H.~Abed, H.~O.~Wang, and R.~C.~Chen, 
	Physica D~{\bf 70}, 154 (1994).
\bibitem{Pyragas3}  A.~Kittel, J.~Parisi, and K.~Pyragas, 
	Phys. Lett. A~{\bf 198}, 433 (1995),
	K.~Pyragas (unpublished).
\bibitem{Pyragas2}  K.~Pyragas and A.~Tama\u{s}evi\u{c}ius, 
	Phys. Lett. A~{\bf 180}, 99 (1993).
\bibitem{Bielawski}  S.~Bielawski, D.~Derozier, and P.~Glorieux, 
	Phys. Rev. E~{\bf 49}, R971 (1994).
\bibitem{SG}  D.~W.~Sukow, D.~J.~Gauthier, M.~E.~Bleich, and J.~E.~S.~Socolar, 
	unpublished.
\bibitem{Hale}  J.~K.~Hale and S.~M.~Verduyn Lunel, 
	{\it Introduction to Functional Differential Equations}, Springer-Verlag 1993.
\bibitem{Ortega} R.~Ortega, 
	Syst. Control Lett. {\bf 8}, 23 (1986).
\bibitem{time-ordering} The time-ordering operator simply means that
	products of matrices 
	occuring in the series expansion of the exponential 
	in Eqn.~(\ref{top}) are to be permuted
	such that the times at which they are evaluated increases monotonically
	from right to left.  The exponential results from the combinatoric factors
	that arise in re-expressing the iterative series expansion for the
	solution of Eqn.~(\ref{Ueq}).  
\bibitem{roots}  Equation (\ref{def}) defines an infinite number of $\mu$
	because of the essential singularity at $\mu=0$.  Physically, this
	arises from the infinite degrees of freedom inherent to time-delayed
	systems, such as Eqn.~(\ref{sys}).
\bibitem{Henrici} See, for example, P.~Henrici, 
	``Applied and Computational Complex Analysis: Volume I''
	(John Wiley \& Sons, New York), section 4.10.
\bibitem{autonomous-note}  In the case of autonomous systems, there is
	always a Floquet multiplier of exactly unity associated with
	an overall time-translation of the system. 
	Since this root does not represent an instability, one
	should work with the function $\frac{g(\mu^{-1})}{1-\mu^{-1}}$
	rather than $g(\mu^{-1})$ itself. 
\bibitem{pre} We note that Eqn.~(\ref{periodicRE}) has a form that has been treated
	in some detail in the control theory literature.  
	The question of stabilizability
	leads to an analysis of the periodic Riccati equation.
	See, e.g., ``The Riccati Equation''
	(S. Bittanti, A. J. Laub, and J. C. Willems, Eds.,
	Springer-Verlag Berlin, Heidelberg 1991) pp. 127-162.
	These treatments are not directly applicable to our problem, however,
	because they assume that the feedback matrix
	can be chosen to be {\em any} periodic function of time.
	In the present case, the time-dependence of ${\bf M}$ is
	highly restricted ---  
	though we are free to choose the unit vector ${\bf\hat{n}}$ and the
	gain $\gamma$, the time-dependence enters only through 
	$\frac{\partial {\bf f}}{\partial \kappa}$ and is determined by the system.
	
\end{references}
\end{document}